\documentclass[twocolumn,english,aps,pra,eqsecnum]{revtex4-1}
\usepackage{mathptmx}
\usepackage{helvet}
\usepackage{courier}
\usepackage[T1]{fontenc}
\usepackage[latin1]{inputenc}
\usepackage{bm}
\usepackage{amsmath}
\usepackage{amssymb}
\usepackage{graphicx}
\usepackage{esint}

\makeatletter

\providecommand{\tabularnewline}{\\}

\@ifundefined{textcolor}{}
{%
 \definecolor{BLACK}{gray}{0}
 \definecolor{WHITE}{gray}{1}
 \definecolor{RED}{rgb}{1,0,0}
 \definecolor{GREEN}{rgb}{0,1,0}
 \definecolor{BLUE}{rgb}{0,0,1}
 \definecolor{CYAN}{cmyk}{1,0,0,0}
 \definecolor{MAGENTA}{cmyk}{0,1,0,0}
 \definecolor{YELLOW}{cmyk}{0,0,1,0}
 }

\makeatother

\usepackage{babel}
\begin{document}

\title{Linear entropy in quantum phase space}

\author{Laura E. C. Rosales-Zárate and P. D. Drummond}

\affiliation{Centre for Atom Optics and Ultrafast Spectroscopy, Swinburne University
of Technology, Melbourne 3122, Australia}
\begin{abstract}
We calculate the quantum Renyi entropy in a phase space representation
for either fermions or bosons. This can also be used to calculate
purity and fidelity, or the entanglement between two systems. We show
that it is possible to calculate the entropy from sampled phase space
distributions in normally ordered representations, although this is
not possible for all quantum states. We give an example of the use
of this method in an exactly soluble thermal case. The quantum entropy
cannot be calculated at all using sampling methods in classical symmetric
(Wigner) or antinormally ordered (Husimi) phase spaces, due to inner
product divergences. The preferred method is to use generalized Gaussian
phase space methods, which utilize a distribution over stochastic
Green's functions. We illustrate this approach by calculating the
reduced entropy and entanglement of bosonic or fermionic modes coupled
to a time-evolving, non-Markovian reservoir.
\end{abstract}
\maketitle

\section{Introduction}

Quantum dynamics and thermal equilibrium states in large many-body
systems have been widely investigated using phase space representations\cite{wigner_32,husimi_40,glauber_63,drummond_gardiner_1980}.
Applications of these methods\cite{Hillery1984} include EPR correlations
in parametric amplifiers, quantum soliton propagation\cite{carter_drummond_87},
non-equilibrium quantum criticality\cite{Dechoum:2004}, quantum dynamics
of Bose-Einstein condensates in two\cite{drummond_corney_99} and
three dimensions\cite{deuar_drummond_07,Ogren2009a}, molecular downconversion\cite{Ogren2009}
and many other problems. Gaussian operator expansions\cite{Corney2004a,Corney2006,corney_drummond_06c,Corney:2003}
are an important extension to these phase space mappings. Unlike earlier
methods, Gaussian phase-space methods can be applied to either fermionic
or bosonic quantum many-body systems\cite{Rahav,Palmieri}. These
methods employ a representation as a distribution over stochastic
Green's functions. They have the essential property that they are
probabilistic, allowing them to scale to large sized systems without
the exponential complexity of the usual orthogonal basis-set methods.
Such methods have proved useful for treating the ground state of the
fermionic Hubbard model\cite{Assaad2005a,Imada_JPS_2007,Corboz_2008},
and dynamical cases where quantum Monte Carlo techniques are impractical\cite{Ogren2010,Ogren2011}.

Yet one of the most fundamental properties of quantum systems, entropy\cite{Neumann},
has not been treated using this approach. In ultra-cold atomic physics
it is usually the entropy that is directly measurable, rather than
the temperature\cite{Thomas_entropy_2007}. This is because there
is no traditional thermal reservoir, so that to measure thermal effects,
it often necessary to use an entropy-conserving adiabatic passage
to a state of known entropy. While some QMC methods can calculate
entropy\cite{Melko_2010_RenyiE_QMCS}, standard phase-space simulations
have not yet done this. Similarly, in quantum information, entropic
concepts like entanglement of formation \cite{BennettEntanglementPhysRevA.53.2046}
and the quantum discord \cite{OlivierQuantumDiscordPhysRevLett.88.017901}
are important measures of quantum behavior. This leads to the question:
can phase-space methods be used to calculate entropy, which is not
a typical quantum observable?

In this paper we investigate quantum entropy calculations using phase-space
representations. Such calculations can be used, for example, to determine
thermodynamics and entanglement of a many-body system\cite{Entanglement}
or to calculate the fidelity of a quantum memory \cite{HeQuantumMemoryPhysRevA.79.022310}.
Alternatively, they can be utilized to assess how close a calculated
state is to a known state, or to check that unitary evolution preserves
the entropy. This is essential in computationally demanding problems,
where it is important to be able to check the validity of a given
quantum simulation. More generally, such investigations throw light
on one of the fundamental problems of quantum statistical physics.
This is the well-known paradox that unitary evolution leaves the quantum
entropy unchanged, while apparently introducing disorder through collisions
and mixing. 

For a probabilistically sampled distributions, we show that the preferred
method for entropy calculations is a Gaussian phase-space representation.
In this approach, the operator basis is comprised of Gaussian operators,
and the corresponding variance or Green's function is used as a stochastic
phase-space variable \cite{Corney2004a}. For ease of computation
we treat the simplest entropy measure, the Renyi or linear entropy\cite{Renyi_1960,Renyi_book}.
An essential ingredient in sampled linear entropy calculations is
a knowledge of the inner products of the generating operators of the
phase space representation. We calculate these inner products for
phase-space representations of either bosonic \cite{Corney:2003}or
fermionic \cite{Corney2006} fields. In the case of fermionic operators,
the results come from an elegant application of Grassmann integration
in a Grassmann space. 

This method is directly useful for calculating a coarse-grained or
localized entropy \cite{Pattanayak}. Coarse-grained entropy can increase
even when the fundamental quantum entropy is invariant, and is a fundamental
entanglement measure. Coarse-graining is then carried out simply by
restricting or projecting the stochastic Greens function onto a subspace
of system modes. As physical applications of these methods, we calculate
the coarse-grained partial entropy for systems of either bosonic or
fermionic modes that are coupled to a general, non-Markovian reservoir.
These cases cannot be treated using master equation methods, and have
applications to many current nanoscale systems in quantum information.
These can be readily solved for the entanglement between the system
and reservoir using the phase-space entropy approach.

By comparison, the commonly used Wigner\cite{wigner_32} and Husimi
Q-function \cite{husimi_40} expansions are not directly useful in
entropy calculations. In the Q-function case, the inner products of
the generating operators are not defined. In the Wigner distribution
case, the entropy obtained from a sampled phase-space distribution
is singular. For normally-ordered distributions, like the Glauber-Sudarshan
P-distribution\cite{glauber_63,sudarshan_63}, the sampled quantum
entropy is well-behaved, but the distribution is singular, except
for classical-like states. This rules out the use of traditional,
classical phase-space mappings for entropy calculations using probabilistic
sampling. The positive P-distribution\cite{drummond_gardiner_1980,gilchrist_gardiner_97,Schack_44},
defined on a double-dimension phase-space, has a well-defined positive
distribution with non-singular inner products. Even in this case,
entropy calculations are nontrivial since, for some quantum states,
the sampling error for entropy calculations diverges.

This paper is organized as follows: In the next section, we describe
the definition of quantum entropy and also describe the phase-space
representations that we will treat in this paper. We also give a general
expression of entropy in terms of phase space-representations and
discuss the evaluation of the sampled entropy. In section III we discuss
coarse graining and reduced entropy and give their expressions in
terms of phase-space projections. In section IV we describe sampled
entropy calculations using fixed variance phase space methods, for
instance the general Cahill-Glauber distribution, as well as the Husimi,
Wigner, Glauber-Sudarshan and positive-P representations. In section
V and VI we discuss general Gaussian phase-space representations for
bosons and fermions respectively. The evaluation of inner products
of Gaussian operators is presented, which allows us to evaluate the
linear entropy and the reduced coarse grained entropy. We evaluate
the linear entropy analytically for thermal states and a comparison
with the sampled entropy using the Glauber-Sudarshan representation
is made. Finally, as physically relevant examples, we evaluate the
coarse grained entropy for both bosons and fermions, in the important
case of a system of particles linearly coupled to a non-Markovian
reservoir. Section VII gives a summary of our results and conclusions.

\section{Entropy and operator representations}

Entropy is a conserved quantity for unitary evolution in quantum mechanics.
Intuitively, entropy \cite{Wehrl,Neumann} is a measure of loss of
information about a physical system. It is invariant for unitary quantum
evolution, and only changes when one considers a subsystem coupled
to a reservoir. Physically, quantum pure state evolution involves
no intrinsic information loss. Of course, this is somewhat counter-intuitive.
One might expect the mixing effects of nonlinear evolution to reduce
information. But this is only true if a restricted or `coarse-grained'
set of measurements is used; in principle, there is no information
loss in pure state evolution, and information or entropy should be
invariant in any simulation of unitary quantum dynamics. Thus, a pure
state will remain a pure state, implying that state purity is a completely
general benchmark for the accuracy of a quantum simulation.

\subsection{Quantum entropy}

Quantum entropy is most commonly defined using the von Neumann\cite{Neumann}
or Shannon entropy\cite{Shannon_1948}:

\begin{equation}
S=-Tr\left(\widehat{\rho}\ln\widehat{\rho}\right)\,\,.
\end{equation}

This is an important physical quantity, related to both information
content and to thermodynamic behavior. However, there are many other
conserved entropic quantities. The existence of these can be thought
of as related to general information conservation. These are recently
discussed \cite{Manfredi} in relation to the Wigner function, where
it is pointed out that any quantity like $S_{F}=Tr\left(F(\widehat{\rho})\right)$
is also conserved. In particular, the linear entropy \cite{Zurek},
which we normalize following Renyi\cite{Renyi_1960}, is defined as:
\begin{equation}
S_{2}=-\ln Tr\left(\widehat{\rho}^{2}\right)\,\,.\label{eq:quantu-Renyi}
\end{equation}
 This has similar properties to the entropy, and measures state purity,
since $S_{2}=0$ for a pure state, while $S_{2}>0$ for a mixed state.
In this note we focus on the linear entropy, which is simplest to
calculate using phase-space methods. This is also true for the fidelity
of $\widehat{\rho}$ to a fiduciary state $\widehat{\rho}_{0}$, 
\begin{equation}
F=Tr\left(\widehat{\rho}\widehat{\rho}_{0}\right)\,\,,
\end{equation}
which is a closely related concept. Such fidelity measures\cite{Amico_2008_RMP}
are useful in evaluating the accuracy of information storage in a
quantum network\cite{Gisin_2002_RMP}, quantum computer\cite{Cirac_1999_QC},
or quantum memory\cite{Gorin_2006}.

The most general class of entropies normally studied in this way are
the general Renyi entropies\cite{Renyi_1960,Renyi_book}, defined
for $p>1$ as: 
\begin{equation}
S_{p}=\frac{1}{1-p}\ln Tr\left(\widehat{\rho}^{p}\right)\,\,.
\end{equation}
 It is known that $S=\lim_{p\rightarrow1}S_{p}$, so the conventional
von Neumann or Shannon entropy can be regained from the general Renyi
entropy in the appropriate limit.

\subsection{Phase-space representations}

A general phase-space representation can be written as\cite{Chaturvedi:1977a,drummond_gardiner_1980}:
\begin{equation}
\widehat{\rho}=\int P(\bm{\lambda})\widehat{\Lambda}(\bm{\lambda})d\bm{\lambda}\,\,,\label{eq:dmatrix}
\end{equation}
 where $P(\bm{\lambda})$ is the probability density over the phase-space,
$\bm{\lambda}$ is a real or complex vector parameter in a general
phase-space, $d\bm{\lambda}$ is the integration measure, and $\widehat{\Lambda}(\bm{\lambda})$
is the representation kernel or operator basis. For simplicity, we
exclude phase-spaces that involve Grassmannian degrees of freedom\cite{Cahill1999a,Olsen_2001}.
These can be extremely useful in analytic calculations, but are not
readily sampled computationally, since the vector parameter $\bm{\lambda}$
is not a real or complex vector.

We consider a bosonic or fermionic quantum field theory with an $M$-dimensional
set of mode operators $\hat{\bm{a}}^{\dagger}\equiv\left[\hat{a}_{1}^{\dagger},\hat{a}_{2}^{\dagger},\ldots\hat{a}_{M}^{\dagger}\right]$.
In the bosonic case, we can define $\delta\hat{\bm{a}}=\hat{\bm{a}}-\bm{\alpha}$
and $\delta\hat{\bm{a}}^{\dagger}=\hat{\bm{a}}^{\dagger}-\bm{\beta}^{\dagger}$
as operator displacements, where in general $\bm{\alpha}$ and $\bm{\beta}^{\dagger}$
are independent complex vectors. In the fermionic case we set these
displacements to zero. The annihilation and creation operators satisfy
(anti) commutation relations, with ($+$) for fermions and ($-$)
for bosons:

\begin{equation}
\left[\hat{a}_{i,}\hat{a}_{j}^{\dagger}\right]_{\pm}=\delta_{ij}.\label{eq:commutators}
\end{equation}

The phase-space representations we will treat in this paper use a
general number-conserving Gaussian operator basis\cite{Corney:2003,Corney2006,corney_drummond_06c},
in which any density matrix $\hat{\rho}$ is expanded in terms of
a basis of Gaussian operators, defined as exponentials of quadratic
operator forms $\widehat{\Lambda}(\bm{\lambda})$, where: 
\begin{equation}
\widehat{\Lambda}(\bm{\lambda})=\frac{1}{{\cal N}}\hat{\Lambda}_{u}\left(\bm{\lambda}\right)=\frac{1}{{\cal N}}:\exp\left[-\delta\hat{\bm{a}}^{\dagger}\underline{\bm{\mu}}\delta\hat{\bm{a}}\right]:
\end{equation}
 Here, $\underline{\bm{\mu}}$ is a complex $M\times M$ matrix so
that $\bm{\lambda}=\left[\bm{\alpha},\bm{\beta}^{\dagger},\underline{\bm{\mu}}\right]$
, ${\cal N}=Tr\left[\widehat{\Lambda}_{u}(\bm{\lambda})\right]$ is
a normalizing factor, and $:\,:$ indicates normal ordering. The normalizing
factor has two forms, for bosons and fermions respectively: 
\begin{eqnarray}
{\cal N}_{b} & = & \det\left[\underline{\bm{\mu}}\right]^{-1}\nonumber \\
{\cal N}_{f} & = & \det\left[2\underline{\bm{I}}-\underline{\bm{\mu}}\right]\,.
\end{eqnarray}

The interpretation as a stochastic Green's function comes from the
identification that $\underline{\bm{\mu}}$ is closely related to
a correlation function of each basis member $\widehat{\Lambda}(\bm{\lambda})$:
\begin{eqnarray}
\underline{\bm{n}}_{b} & = & \underline{\bm{\mu}}^{-T}-\underline{\bm{I}}\nonumber \\
\underline{\bm{n}}_{f} & = & \left[2\underline{\bm{I}}-\underline{\bm{\mu}}\right]^{-T}\,.\label{eq:Green_sf}
\end{eqnarray}
In either case, the stochastic average of $\underline{\bm{n}}$ over
the distribution $P$ is physically a normally ordered many-body Green's
function, so that: 
\begin{equation}
\left\langle \hat{a}_{i}^{\dagger}\hat{a}_{j}\right\rangle =\left\langle n_{ij}+\beta_{i}^{*}\alpha_{j}\right\rangle _{P}\,.
\end{equation}

In traditional, classical types of phase-space - for example, the
Wigner function approach - the random variable or phase-space coordinate
is a stochastic position or momentum. In the case of a general Gaussian
phase-space, the random variable is a stochastic correlation function. 

We see immediately, from Eq. (\ref{eq:quantu-Renyi}), that the Renyi
entropy in a phase-space representation is: 
\begin{equation}
S_{2}=-\ln\iint P(\bm{\lambda})P(\bm{\lambda}')Tr\left(\widehat{\Lambda}(\bm{\lambda})\widehat{\Lambda}(\bm{\lambda}')\right)\, d\bm{\lambda}d\bm{\lambda}'\,.
\end{equation}

The evaluation of inner products of Gaussian operators of form $Tr\left(\widehat{\Lambda}(\bm{\lambda})\widehat{\Lambda}(\bm{\lambda}')\right)$
is therefore a central task in calculations of linear entropy using
phase-space representations.

\subsection{Sampled entropy}

For computational purposes, distributions always exist such that $P(\bm{\lambda})$
has positive values, and it can be interpreted as a probability in
these cases. One can then sample the distribution $ $N times to obtain
a sampled estimate $\widehat{\rho}_{S}$, such that: 
\begin{equation}
\widehat{\rho}\approx\widehat{\rho}_{S}=\frac{1}{N}\sum_{j=1}^{N}\widehat{\Lambda}(\bm{\lambda}_{j})\,\,.
\end{equation}
 This approximation becomes an exact equality in the limit of $N\rightarrow\infty$,
provided the sampling is unbiased. Given a set of samples $\bm{\lambda}_{i}$,
we can now calculate the linear entropy as follows: 
\begin{eqnarray}
S_{2} & \approx & -\ln\left[\frac{1}{N^{2}}\sum_{i,j=1}^{N}Tr\left(\widehat{\Lambda}(\bm{\lambda}_{i})\widehat{\Lambda}(\bm{\lambda}_{j}')\right)\right]\,\,.\label{eq:S2_sampled}
\end{eqnarray}

This, however, requires a double sampling of the population. In phase-space
representations the kernel of the representation consists of non-orthogonal
operators, so that the operator inner product $Tr\left(\widehat{\Lambda}(\bm{\lambda}_{i})\widehat{\Lambda}(\bm{\lambda}{}_{j})\right)$
is non-vanishing even if $\bm{\lambda}_{i}\neq\bm{\lambda}_{j}$.
It is desirable that the two sets of samples $\bm{\lambda}_{i},\bm{\lambda}_{j}'$
are independent of each other, to prevent sampling biases. The above
result has obvious extensions to other entropies. For example, the
general Renyi entropy involves a $p$-fold summation over sampling
indices: 
\begin{eqnarray}
S_{p} & \approx & \frac{1}{1-p}\ln\left[\frac{1}{N^{p}}\sum_{\mathbf{j}=1}^{N}Tr\left(\widehat{\Lambda}(\bm{\lambda}_{j_{1}})\ldots\widehat{\Lambda}(\bm{\lambda}"_{j_{p}})\right)\right]\,,
\end{eqnarray}
 but clearly the linear entropy is computationally the simplest. In
the remainder of this paper, we focus on the question of how to evaluate
the inner-products of the Gaussian phase-space basis set, which is
the essential ingredient in calculating a linear entropy or fidelity
measure, and how to apply this in physically relevant situations.
We note that for some calculations it is useful to allow $P(\bm{\lambda})$
to have complex values \cite{DrummondGardinerWallsPhysRevA.24.914},
in which cases the entropy is best calculated analytically.

\section{Coarse graining and reduced entropy}

There is a fundamental paradox in understanding quantum entropy. For
an isolated quantum system, all of the entropies defined above are
invariant under unitary evolution, even including particle-particle
interactions. This appears to defy conventional wisdom, which is that
for a many-body system the effect of particle collisions is to cause
mixing, and hence increase disorder. Thus, collisions appear to increase
the entropy. Such expectations contradict the entropy invariance of
unitary evolution, which is at the heart of such famous controversies
as the Bekenstein-Hawking black hole information loss paradox \cite{BekensteinPhysRevD.9.3292,HawkingPhysRevD.14.2460}. 

However, these paradoxes are easily resolved at a practical level.
Typically, in many experiments only part of the density matrix is
measurable. For example, one may only have experimental access to
measurements of the low momentum modes. Under these conditions, one
can separate the Hilbert space into a measured part $\mathcal{H}_{A}$
and unmeasured part $\mathcal{H}_{B}$, so that the entire Hilbert
space is $\mathcal{H}=\mathcal{H}_{A}\otimes\mathcal{H}_{B}$. Other
separations of measured and unmeasured operations are also possible,
using the method of communication alphabets \cite{CavesDrummondRevModPhys.66.481}.
An interesting recent proposal of this type is to employ the many-body
energy eigenstates as a communication alphabet to define entropy \cite{PolkolnikovEntropy}. 

Here, for definiteness, we focus on a division of the Hilbert space
into measured and unmeasured single-particle modes. These could, for
example, correspond to a physical partition into distinct spatial
locations. The two parts of the quantum wave-function then become
entangled during time-evolution under a Hamiltonian that couples the
two parts. This means that part of the quantum information is only
accessible through measurement of correlations. An estimate of this
relative entropy \cite{VedralRelativeEntropyRevModPhys.74.197} based
on measurements reveals an apparent increase in entropy, or loss of
information due to entanglement \cite{BennettEntanglementPhysRevA.53.2046}.

If we trace out the unmeasured part of Hilbert space, denoting this
trace over $\mathcal{H}_{B}$ as $Tr_{B}$, we obtain the reduced
density matrix that corresponds to operational measurement on $A$:
\begin{equation}
\hat{\rho}_{A}=Tr_{B}\left[\hat{\rho}\right]\,.
\end{equation}
Such a reduced density matrix can experience increased entropy - called
entanglement entropy - even when the total entropy is conserved. The
corresponding reduced entropy is then:
\begin{equation}
S_{p}^{red}=\frac{1}{1-p}\ln Tr_{A}\left(\widehat{\rho}_{A}^{p}\right)\,\,.\label{eq:red_entropy}
\end{equation}

This reduced entropy is an important measure of quantum entanglement.
In the case of a pure state, $S_{p}^{red}>0$ is both necessary and
sufficient for entanglement. This can also be extended to the case
of mixed states. In this case, one must generalize the approach, to
take account the possibility that the original state was a mixed state
\cite{BennettEntanglementPhysRevA.53.2046}.

\subsection{Phase-space projections}

In the case of Gaussian phase-space expansions, all our entropy results
are also applicable to the reduced entropy, in which case we must
replace the phase-space basis $\widehat{\Lambda}(\bm{\lambda})$ by:
\begin{equation}
\widehat{\Lambda}^{A}(\bm{\lambda})=Tr_{B}\left[\widehat{\Lambda}(\bm{\lambda})\right]\label{eq:ReducedBasis}
\end{equation}
With such a replacement, the trace used in the following calculations
must be replaced by a reduced trace over $\mathcal{H}_{A}$ for consistency.
If coarse-graining is carried out on a modal basis, we can divide
up the modes into two sets: $\hat{\bm{a}}\equiv\left[\hat{a}_{1},\hat{a}_{2},\ldots\hat{a}_{M}\right]\equiv\left[\hat{\bm{a}}^{A},\hat{\bm{a}}^{B}\right]$.
Here the modes $\hat{\bm{a}}^{A}$ may comprise only low-momentum
modes, or alternatively, only modes localized to part of an apparatus. 

In such cases, $\widehat{\Lambda}^{A}(\bm{\lambda})$ depends on a
new set of parameters $\bm{\lambda}^{A}\equiv(\bm{\alpha}^{A},\bm{\beta}^{A\dagger},\underline{\bm{\mu}}^{A})$.
The reduced displacements are just the projection of the full displacements
onto the reduced Hilbert space, while the reduced covariance can be
evaluated using standard trace identities. We first write the original
matrix $\underline{\bm{\mu}}$ in a block form as: 
\begin{equation}
\underline{\bm{\mu}}=\left[\begin{array}{cc}
\underline{\bm{\mu}}^{AA} & \underline{\bm{\mu}}^{AB}\\
\underline{\bm{\mu}}^{BA} & \underline{\bm{\mu}}^{BB}
\end{array}\right]\,,
\end{equation}
so that the Gaussian exponent term becomes: 
\begin{equation}
\delta\hat{\bm{a}}^{\dagger}\underline{\bm{\mu}}\delta\hat{\bm{a}}=\delta\hat{\bm{a}}^{A\dagger}\underline{\bm{\mu}}^{AA}\delta\hat{\bm{a}}^{A}+\delta\hat{\bm{a}}^{A\dagger}\underline{\bm{\mu}}^{AB}\delta\hat{\bm{a}}^{B}+(A\leftrightarrow B)
\end{equation}

Next, the relevant traces over the unobserved subspace $B$ are evaluated
using coherent state identities: 
\begin{eqnarray}
Tr_{b}[\hat{O}] & = & \frac{1}{\pi^{M}}\int d^{2M}\bm{\alpha}\langle\bm{\alpha}\vert\hat{O}\vert\bm{\alpha}\rangle,\nonumber \\
Tr_{f}[\hat{O}] & = & \int d^{2M}\bm{\alpha}\langle-\bm{\alpha}\vert\hat{O}\vert\bm{\alpha}\rangle\label{eq:TraceIdentity}
\end{eqnarray}

This gives the result that the reduced basis set remains Gaussian,
but with a modified covariance: 
\begin{equation}
\widehat{\Lambda}^{A}(\bm{\lambda})=\widehat{\Lambda}^{A}(\bm{\alpha}^{A},\bm{\beta}^{A\dagger},\underline{\bm{\mu}}^{A})
\end{equation}
where the reduced covariance matrix $\underline{\bm{\mu}}^{A}$ is
given, for bosons and fermions respectively, by:

\begin{eqnarray}
\underline{\bm{\mu}}_{b}^{A} & = & \underline{\bm{\mu}}^{AA}-\underline{\bm{\mu}}^{AB}\left[\underline{\bm{\mu}}^{BB}\right]^{-1}\underline{\bm{\mu}}^{BA}\nonumber \\
\underline{\bm{\mu}}_{f}^{A} & = & \underline{\bm{\mu}}^{AA}+\underline{\bm{\mu}}^{AB}\left[2\underline{\bm{I}}-\underline{\bm{\mu}}^{BB}\right]^{-1}\underline{\bm{\mu}}^{BA}.\label{eq:reducedmatrix}
\end{eqnarray}

The important result here is that for a Gaussian basis, coarse-graining
via mode-projection leaves the phase-space representation invariant.
Just as for the full Gaussian expansion, there is a reduced Green's
function for these Gaussian operators in the subspace. From Eq. (\ref{eq:Green_sf})
in the previous section, this must have the standard form of:
\begin{eqnarray}
\underline{\bm{n}}_{b}^{A} & = & \left[\underline{\bm{\mu}}^{A}\right]^{-T}-\underline{\bm{I}}^{A}\nonumber \\
\underline{\bm{n}}_{f}^{A} & = & \left[2\underline{\bm{I}}^{A}-\underline{\bm{\mu}}^{A}\right]^{-T}\,.\label{eq:Green_sf-1}
\end{eqnarray}

The results for the reduced stochastic Green's function for bosons
and fermions can now be written, using standard matrix block reduction
algebra, in terms of the block representation of the original stochastic
matrix $\underline{\bm{n}}$, which also has a decomposition:
\begin{equation}
\underline{\bm{n}}=\left[\begin{array}{cc}
\underline{\bm{n}}^{AA} & \underline{\bm{n}}^{AB}\\
\underline{\bm{n}}^{BA} & \underline{\bm{n}}^{BB}
\end{array}\right]\,,
\end{equation}

We find that the trace reduction simply gives the diagonal block in
the first quadrant. 

\begin{eqnarray}
\underline{\bm{n}}_{b}^{A} & = & \underline{\bm{n}}_{b}^{AA}\nonumber \\
\underline{\bm{n}}_{f}^{A} & = & \underline{\bm{n}}_{f}^{AA}.
\end{eqnarray}
This has a simple physical explanation. We naturally expect that any
correlation function that is restricted just to the $A$ Hilbert space
will have no dependence on measurable correlations of the B Hilbert
space. This physical property of the full Green's function also holds
for the stochastic Green's functions as well.

The basis is mapped to new values, and the reduced density matrix
$\widehat{\rho}^{A}=Tr_{B}[\hat{\rho}]$ of Eq.$\ $(\ref{eq:red_entropy})
can be written in the reduced Gaussian representation in terms of
the reduced Green's function as:
\begin{eqnarray}
\widehat{\rho}^{A} & = & Tr_{B}\int P(\bm{\underline{n}})\Lambda(\underline{\bm{n}})d\underline{\bm{n}}\nonumber \\
 & = & \int P(\underline{\bm{n}})\Lambda^{A}(\underline{\bm{n}}^{A})d\underline{\bm{n}}
\end{eqnarray}

Next, we can introduce the corresponding reduced distribution function:
\begin{equation}
P^{A}(\underline{\bm{n}}^{A})\equiv\left[\int P(\underline{\bm{n}})d\underline{\bm{n}}^{\backslash A}\right]
\end{equation}
where $\underline{\bm{n}}^{\backslash A}$ is the relative complement
of $\underline{\bm{n}}^{A}$, i.e., the set of all variables in $\underline{\bm{n}}$
that are not included in $\underline{\bm{n}}^{A}$.

$ $ Using the definition of the reduced density matrix, the linear
coarse grained entropy, Eq. $\ $(\ref{eq:red_entropy}), for $p=2$,
is:

\begin{equation}
S_{2}^{red}=-\ln\iint P^{A}(\underline{\bm{n}}^{A})P^{A}(\underline{\bm{n}}'^{A})Tr_{A}\left(\widehat{\Lambda}^{A}(\underline{\bm{n}}^{A})\widehat{\Lambda}^{A}(\underline{\bm{n}}'^{A})\right)\, d\underline{\bm{n}}^{A}d\underline{\bm{n}}'^{A}.\label{eq:S2_red}
\end{equation}

\section{Fixed variance phase-spaces}

To evaluate the entropy from a set of phase-space samples, we need
the inner-product of the kernel members. This depends on how the phase-space
is parametrized, either through changing the displacement, or the
variance, or both. Traditional phase-spaces for bosons utilize a displacement-based
approach, which is the most similar to classical phase-space ideas.
In the case of fermions, the displacements must be Grassmann variables,
not c-numbers \cite{Cahill1999a}, which means that only the \emph{variances}
can be readily sampled computationally. In this section, we treat
fixed variance phase-spaces, which are therefore bosonic.

\subsection{Cahill-Glauber phase-space}

The traditional mappings of bosonic fields to a classical phase-space
utilize a single classical displacement. These can all be written
in a unified form as \cite{Corney:2003}: 
\begin{equation}
\widehat{\Lambda}_{s}(\bm{\lambda})=\frac{1}{{\cal N}}:\exp\left[-(\widehat{\bm{a}}^{\dagger}-\bm{\alpha}^{\dagger})\underline{\bm{\mu}}(\widehat{\bm{a}}-\bm{\alpha})\right]:
\end{equation}

Here $\bm{\alpha},$$\bm{\alpha}^{\dagger}$ are a complex vectors,
and $\underline{\bm{\mu}}$ is held constant so that $\bm{\lambda}=\bm{\alpha}$.
There are three famous cases, corresponding to different values of
$\underline{\bm{\mu}}=2\underline{\bm{I}}/\left(s+1\right)$, where
$s=0,\pm1$. Cahill and Glauber \cite{CG-Q} have calculated the inner
product for these $s$-ordered representations, which includes the
diagonal P-representation ($s=1$) and the Wigner representation ($s=0$),
as special cases. Their results are that, for $s>0$: 
\begin{eqnarray}
Tr\left(\widehat{\Lambda}_{s}(\bm{\alpha})\widehat{\Lambda}_{s}(\bm{\alpha}')\right) & = & \frac{1}{s}\exp\left[-\left|\bm{\alpha}-\bm{\alpha}'\right|^{2}/s\right]
\end{eqnarray}

We note that the Husimi representation with $s=-1$ has no well-defined
inner-product for its basis set members, as the product trace is divergent.
Thus, a point-sampled Q-function is not a useful way to calculate
the entropy, without additional assumptions. More sophisticated techniques
would be needed in this case. One could, for example, expand the Q-function
using Gaussian wavelets, instead of delta-functions, so that the sampling
expansion uses smoother functions. However, since different types
of s-ordering are interrelated through Gaussian convolutions, this
simply generates another member of the class of Gaussian operator
expansions.

\subsection{Wigner representation}

Strictly speaking, the only positive Wigner distributions are the
Gaussian ones that represent certain special cases, including thermal,
coherent and squeezed states. Nevertheless, one often wishes to use
a truncated Wigner time-evolution equation, which generates positive
Wigner distributions as an approximation to the full time-evolution.
This has a close analogy with a classical phase-space, for which entropy
can also be calculated in the classical sense.

One can treat the Wigner case as the Cahill-Glauber representation
in the limit of $s\rightarrow0$, where:

\begin{eqnarray}
Tr\left(\widehat{\Lambda}_{0}(\bm{\alpha})\widehat{\Lambda}_{0}(\bm{\alpha}')\right) & = & \pi^{M}\delta^{M}\left(\bm{\alpha}-\bm{\alpha}'\right)\,,
\end{eqnarray}
 which is highly singular. As in the Husimi case, point-sampling doesn't
provide a useful estimate of the purity. Two distinct samples will
not have identical points in phase-space, except for points of measure
zero where the samples are equal. This demonstrates the nontrivial
nature of estimating quantum entropy in sampled phase-space representations.
One can understand this from the perspective that the coherent states
are the only pure states with a positive Wigner function. These have
a finite distribution variance, but a zero quantum entropy.

This result is consistent with other calculations. It is known that
one can estimate $S_{2}$ in a Wigner representation through \cite{Manfredi}
: 
\begin{equation}
S_{2}=-\ln\pi^{M}\int W^{2}\left(\bm{\alpha}\right)d^{2M}\bm{\alpha}\,,
\end{equation}
 which is identical with the delta-correlated trace result given above.
This can be used when $W$ is known analytically, but it is not computationally
useful when we only have access to a sampled estimate of $W\left(\bm{\alpha}\right)$.

\subsection{Glauber-Sudarshan}

In the case of the normally-ordered Glauber-Sudarshan representation,
$s=1$. This corresponds to an expansion in coherent-state projectors,
so that $\widehat{\Lambda}_{1}(\bm{\alpha})\equiv\left|\bm{\alpha}\right\rangle \left\langle \bm{\alpha}\right|$,
where $\left|\bm{\alpha}\right\rangle $ is a coherent state, and
\begin{eqnarray}
Tr\left(\widehat{\Lambda}_{1}(\bm{\alpha})\widehat{\Lambda}_{1}(\bm{\alpha}')\right) & = & \exp\left[-\left|\bm{\alpha}-\bm{\alpha}'\right|^{2}\right]
\end{eqnarray}

Here the linear entropy is well-behaved, and both linear coupling
and damping can be treated exactly. However, there is no corresponding
stochastic process in this case, for nonlinear evolution of an interacting
system, and many nonclassical states involve a nonpositive or singular
distribution. For a positive distribution, the only pure states in
this representation are coherent states. Provided a Glauber-Sudarshan
distribution exists, a direct point-sampling is enough to obtain the
entropy. One can easily obtain the entropy of a non-interacting thermal
state, which always has a well-defined Glauber-Sudarshan distribution.
For example, the vacuum state has a delta-function distribution, and
so clearly one has: $Tr\left(\widehat{\Lambda}_{1}(\bm{\alpha})\widehat{\Lambda}_{1}(\bm{\alpha}')\right)=1$
, and hence $S_{2}=S_{p}=0$ as expected.

For a thermal case with: 
\begin{equation}
\hat{\rho}_{th}\propto\exp\left[-\widehat{\bm{a}}^{\dagger}\widehat{\bm{a}}/k_{b}T\right]=:\exp\left[-\widehat{\bm{a}}^{\dagger}\left[1+\underline{\bm{n}}\right]^{-1}\widehat{\bm{a}}\right]:
\end{equation}
where $\underline{\bm{n}}$ is the thermal Bose-Einstein occupation
number, clearly 
\begin{equation}
n_{_{kk'}}\equiv\frac{\delta_{kk'}}{e^{E_{k}/k_{b}T}-1}
\end{equation}
, where $k_{b}$ is the Boltzmann constant. 

\begin{figure}
\includegraphics[scale=0.5]{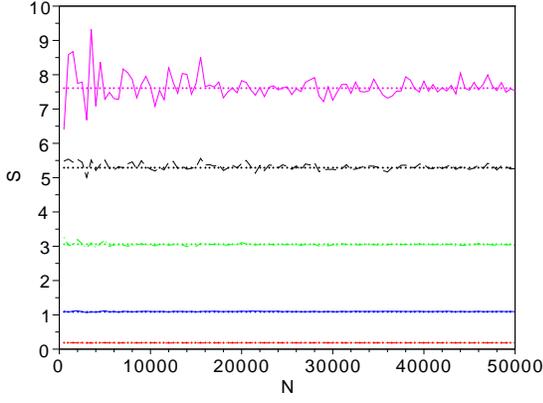}

\caption{Comparison of the linear entropy for thermal states using the Glauber-Sudarshan
representation and the Gaussian representation for bosons. The dotted
line is the exact result using the Gaussian representation for bosons
for $\underline{n}=0.1,\:1,\:10,\:100,\,1000$ , with smallest occupation
numbers having the lowest entropy. Here $N$ is the number of samples
used. \label{Fig:comparison_thermal_entropy}}
\end{figure}

Here one finds that in the Glauber-Sudarshan representation, one has:
\begin{equation}
P\left(\bm{\alpha}\right)=exp\left[-\bm{\alpha}^{\dagger}\underline{\bm{n}}^{-1}\bm{\alpha}'\right].
\end{equation}
Therefore we can use the results of $P(\bm{\alpha})$ in order to
sample the entropy for the thermal states. In Figure (\ref{Fig:comparison_thermal_entropy})
we show the results of the sampled linear entropy as a function of
the number of samples $N$ for different values of the thermal Bose-Einstein
occupation number $\underline{\bm{n}}$, compared with exact results
obtained in the next section.

The generators in this case are coherent state projection operators,
which means that obtaining a coarse-grained entropy is straightforward.
On dividing the modes into two groups, $A$ and $B$, one can simply
write the coherent state as an outer product: 
\[
\left|\bm{\alpha}\right\rangle =\left|\bm{\alpha}^{A}\right\rangle \left|\bm{\alpha}^{B}\right\rangle 
\]
so that:
\[
\widehat{\Lambda}_{1}(\bm{\alpha})=\widehat{\Lambda}_{1}(\bm{\alpha}^{A})\widehat{\Lambda}_{1}(\bm{\alpha}^{B})
\]
where $\widehat{\Lambda}_{1}(\bm{\alpha}^{A})\equiv\left|\bm{\alpha}^{A}\right\rangle \left\langle \bm{\alpha}^{A}\right|$
is a coherent state projector in the reduced Hilbert space.

\subsection{Positive-P representation}

The positive P-representation extends the Glauber-Sudarshan representation
into a space of double the classical dimension, with $\bm{\lambda}=(\bm{\alpha},\bm{\beta})$.
This the advantage that any state or density matrix has a positive
probability expansion. Unlike the Husimi Q-function, the basis set
has a non-singular inner-product, which allows the entropy to be calculated
through sampling techniques. The kernel can be written in an alternate
form as an hermitian projection operator\cite{Chaturvedi:1977a,drummond_gardiner_1980}:
\begin{equation}
\widehat{\Lambda}(\bm{\lambda})=\frac{\left|\bm{\alpha}\right\rangle \left\langle \bm{\beta}\right|}{\left\langle \bm{\beta}\right.\left|\bm{\alpha}\right\rangle }.
\end{equation}
Just as in the Glauber-Sudarshan case, the issue of coarse-graining
is a straightforward one of simply dividing the modes into two groups,
and replacing $\widehat{\Lambda}(\bm{\lambda})$ by its reduced version,
$\widehat{\Lambda}(\bm{\lambda}^{A})$. The inner-product is always
well-defined, being just a Gaussian form in the displacement vectors:
\begin{eqnarray}
Tr\left(\widehat{\Lambda}(\bm{\lambda})\widehat{\Lambda}(\bm{\lambda}')\right) & = & \frac{\left\langle \bm{\beta}\right.\left|\bm{\alpha}'\right\rangle \left\langle \bm{\beta}'\right.\left|\bm{\alpha}\right\rangle }{\left\langle \bm{\beta}\right.\left|\bm{\alpha}\right\rangle \left\langle \bm{\beta}'\right.\left|\bm{\alpha}'\right\rangle }\nonumber \\
 & = & \exp\left[-\left(\bm{\beta}-\bm{\beta}'\right)^{\dagger}\left(\bm{\alpha}-\bm{\alpha}'\right)\right]\,.\label{eq:Innerproduct}
\end{eqnarray}
 In all cases, a highly localized distribution is guaranteed to exist,
from the fundamental existence theorem of the positive-P representation.
This states that at least one canonical, positive distribution $P(\bm{\alpha},\bm{\beta})$
always exists for any $\hat{\rho}$, with: 
\begin{equation}
P(\bm{\alpha},\bm{\beta})=\frac{1}{\left(2\pi\right)^{2M}}e^{-\left|\bm{\alpha}-\bm{\beta}\right|^{2}/4}\left\langle \frac{\bm{\alpha}+\bm{\beta}}{2}\right|\widehat{\rho}\left|\frac{\bm{\alpha}+\bm{\beta}}{2}\right\rangle \label{eq:canonical}
\end{equation}

While this distribution always exists, and is suitable for calculating
moments, it generally leads to large sampling errors when calculating
the entropy. This is due to the fact that when $|\alpha_{i}-\beta_{i}|^{2}\gg1$,
in Eq. (\ref{eq:Innerproduct}), the cross-terms can become exponentially
large, as these are not sufficiently bounded by the exponentials in
the canonical form, Eq. (\ref{eq:canonical}). 

In summary, we see that calculating entropies using a displacement
based phase-space expansion with point sampling is non-trivial. With
traditional phase-space expansions, either the basis has singular
inner-products, or the distribution is non-positive, or both. For
the Glauber-Sudarshan representation of a thermal state, the distribution
is well-behaved and the linear entropy can be computed. In the positive-P
case, a positive distribution always exists, and the basis has nonsingular
inner products. However, even in this case the entropic sampling error
can diverge for nonclassical states.

\section{Gaussian representations for bosons}

An alternative way to represent quantum states in phase-space, is
to use a general representation in terms of Gaussian operators. These
types of phase-spaces can in principle combine the displacement and
variance-based approach. However, for definiteness, in this section
we will treat the case where the representation is entirely variance
based. Such an approach has a clear intuitive meaning. In this approach,
the physical many-body system is treated as a distribution over stochastic
Green's functions, whose average in the observed Green's function
or correlation function. We note that the basis set includes non-hermitian
terms for completeness, which means that the stochastic Green's functions
themselves can be nonhermitian.

\subsection{Un-normalized Gaussians}

For the bosonic case, we must evaluate the trace of the product of
two un-normalized bosonic Gaussian operators, $B\left(\underline{\bm{\mu}},\underline{\bm{\nu}}\right)=Tr\left[\hat{\Lambda}_{u}\left(\underline{\bm{\mu}}\right)\hat{\Lambda}_{u}\left(\underline{\bm{\nu}}\right)\right]$
for the M-mode case:

\begin{equation}
B\left(\underline{\bm{\mu}},\underline{\bm{\nu}}\right)=Tr\left[:e^{-\hat{\bm{a}}^{\dagger}\underline{\bm{\mu}}\hat{\bm{a}}}::e^{-\hat{\bm{a}}^{\dagger}\underline{\bm{\nu}}\hat{\bm{a}}}:\right],
\end{equation}

Using the expressions for the trace of an operator, Eq. (\ref{eq:TraceIdentity}),
and the expansion of the identity operator in terms of the bosonic
coherent states:

\begin{equation}
\frac{1}{\pi^{M}}\int d^{2M}\bm{\alpha}\vert\bm{\alpha}\rangle\langle\bm{\alpha}\vert=\hat{I},
\end{equation}
we obtain: 
\begin{eqnarray}
B\left(\underline{\bm{\mu}},\underline{\bm{\nu}}\right) & = & \frac{1}{\pi^{2M}}\int d^{2M}\bm{\alpha}d^{2M}\bm{\beta}\left\langle \bm{\alpha}\right|:e^{-\hat{\bm{a}}^{\dagger}\underline{\bm{\mu}}\hat{\bm{a}}}:\times\nonumber \\
 &  & \times\left|\bm{\beta}\right\rangle \left\langle \bm{\beta}\right|:e^{-\hat{\bm{a}}^{\dagger}\underline{\bm{\nu}}\hat{\bm{a}}}:\left|\bm{\alpha}\right\rangle .
\end{eqnarray}
 Expanding the normal-ordered exponential, and using the standard
eigenvalue properties for the bosonic coherent states: $\hat{\bm{a}}\left|\bm{\alpha}\right\rangle =\bm{\alpha}\left|\bm{\alpha}\right\rangle $
gives:

\begin{eqnarray}
B\left(\underline{\bm{\mu}},\underline{\bm{\nu}}\right) & = & \frac{1}{\pi^{2M}}\int d^{2M}\bm{\alpha}d^{2M}\bm{\beta}\left\langle \bm{\alpha}\right|e^{-\bm{\alpha}^{\dagger}\underline{\bm{\mu}}\bm{\beta}}\times\nonumber \\
 &  & \left|\bm{\beta}\right\rangle \left\langle \bm{\beta}\right|e^{-\bm{\beta}^{\dagger}\underline{\bm{\nu}}\bm{\alpha}}\left|\bm{\alpha}\right\rangle .
\end{eqnarray}
 From the inner product of two coherent states, we finally obtain
a Gaussian integral over $2M$ complex coordinates: 
\begin{equation}
B\left(\underline{\bm{\mu}},\underline{\bm{\nu}}\right)=\frac{1}{\pi^{2M}}\int d^{2M}\bm{\alpha}d^{2M}\bm{\beta}e^{-\bm{\alpha}^{\dagger}\underline{\bm{\mu}}\bm{\beta}-\bm{\beta}^{\dagger}\underline{\bm{\nu}}\bm{\alpha}-\left|\bm{\alpha}-\bm{\beta}\right|^{2}}.
\end{equation}

Next, introducing a double-dimension vector: 
\begin{equation}
\bm{\gamma}=\left[\begin{array}{c}
\bm{\alpha}\\
\bm{\beta}
\end{array}\right]\,,
\end{equation}
 we can write this as: 
\begin{eqnarray}
B\left(\underline{\bm{\mu}},\underline{\bm{\nu}}\right) & = & \frac{1}{\pi^{2M}}\int d^{4M}\bm{\gamma}e^{-\bm{\gamma}^{\dagger}\underline{\bm{\Gamma}}\bm{\gamma}}.\nonumber \\
 & = & \det\left[\underline{\bm{\Gamma}}\right]^{-1}
\end{eqnarray}
 where we have used the standard identity(\cite{Integrals}) for an
N-dimensional Gaussian complex integrals, and introduced a double-dimension
matrix, 
\begin{equation}
\underline{\bm{\Gamma}}=\left[\begin{array}{cc}
\underline{\bm{I}} & \underline{\bm{\mu}}-\underline{\bm{I}}\\
\underline{\bm{\nu}}-\underline{\bm{I}} & \underline{\bm{I}}
\end{array}\right]
\end{equation}

Therefore, on simplifying the determinant, we obtain: 
\begin{equation}
B\left(\underline{\bm{\mu}},\underline{\bm{\nu}}\right)=\det\left[\bm{\mathrm{I}}\,-\left(\underline{\bm{\mu}}-\underline{\bm{I}}\right)\left(\underline{\bm{\nu}}-\underline{\bm{I}}\right)\right]^{-1}
\end{equation}

\subsection{Normalized Gaussians}

It is useful to rewrite these expressions in terms of the underlying
stochastic Green's functions. These are the normally ordered correlations
of the basis sets, defined so that: 
\begin{equation}
n_{ij}=Tr\left[\widehat{\Lambda}(\mathbf{n})\hat{a}_{i}^{\dagger}\hat{a}_{j}\right]
\end{equation}
 Using this definition, the normalized Gaussian generators are: 
\begin{equation}
\widehat{\Lambda}(\mathbf{n})=\frac{1}{\det\left[\underline{\bm{\mathrm{I}}}+\underline{\mathbf{n}}\right]}:\exp\left[-\hat{\bm{a}}^{\dagger}\left[\underline{\bm{\mathrm{I}}}+\underline{\mathbf{n}}\right]^{-1}\hat{\bm{a}}\right]:
\end{equation}

We note that there is a restriction on the values of $\underline{\mathbf{n}}$
, which is that $\Re\left\{ \underline{\bm{\mathrm{I}}}+\underline{\mathbf{n}}\right\} $
must have positive definite eigenvalues in order for the basis operators
to be normalizable, and hence for the Gaussian generators to be in
the Hilbert space.

Applying this normalization to the results given above, one finds
that: 
\begin{equation}
Tr\left[\widehat{\Lambda}(\mathbf{n})\widehat{\Lambda}(\mathbf{m})\right]=\det\left[\underline{\bm{\mathrm{I}}}+\underline{\mathbf{n}}+\underline{\mathbf{m}}\right]^{-1}\label{eq:inner_product_bosons}
\end{equation}
For Renyi entropy calculations there is another restriction. This
is that all pairs of stochastic samples must have the property that
$\Re\left\{ \underline{\bm{\mathrm{I}}}+\underline{\mathbf{n}}+\underline{\mathbf{m}}\right\} $
has positive definite eigenvalues to calculate the entropy using sampling
methods. Under this restriction, the inner products are well-defined. 

In order to illustrate the technique of the Gaussian representation
for bosons in the evaluation of the linear entropy and coarse grained
entropy respectively, in the next two subsections we will evaluate
the linear entropy of thermal states and the coarse grained entropy
of a system coupled to a non-Markovian reservoir.

\subsection{Thermal linear entropy for bosons}

The linear entropy for thermal states is evaluated as previously,
using Eq.$\ $(\ref{eq:S2_sampled}) and the result of Eq.$\ $(\ref{eq:inner_product_bosons})
for the single-mode case. When the density matrix is thermal, only
a single basis set member is required, and:
\begin{eqnarray}
S_{2} & = & -\ln Tr\left(\widehat{\Lambda}^{2}(\mathbf{n}_{th})\right)\nonumber \\
 & = & \ln\det\left[\underline{\bm{\mathrm{I}}}+2\underline{\mathbf{n}}_{th}\right]
\end{eqnarray}

For the single-mode case we know that the thermal Green's function
is a scalar: $\underline{\mathbf{n}}_{th}=n_{th}$, where $n_{th}$
is the Bose-Einstein occupation number at a given temperature. In
Table 1 we show the results for the linear entropy using the Gaussian
representation for bosons. In Figure (\ref{Fig:comparison_thermal_entropy})
we show the comparison of the results using the Gaussian phase-space
representation, results of Table(\ref{Table:Linear_entropy_GRB}),
and the results from the sampling using the Glauber-Sudarshan representation
as a function of the number of samples, giving excellent agreement
in the limit of large numbers of samples. 

Clearly there is a great improvement in efficiency in this case, relative
to the Glauber-Sudarshan approach. Only one Gaussian phase-space sample
is needed, instead of up to $50000$ samples using more traditional
phase-space methods. 

\begin{table}
\begin{tabular}{|c|c|}
\hline 
$n_{th}$ & $S_{2}$\tabularnewline
\hline 
0.01 & 0.0198\tabularnewline
\hline 
0.1 & 0.1823\tabularnewline
\hline 
1 & 1.0986\tabularnewline
\hline 
10 & 3.0445\tabularnewline
\hline 
100 & 5.3033\tabularnewline
\hline 
1000 & 7.6014\tabularnewline
\hline 
\end{tabular}

\caption{Linear entropy for thermal states using the Gaussian representation
for bosons.}
\label{Table:Linear_entropy_GRB}
\end{table}

\subsection{Coarse-grained entropy for bosons}

We now wish to consider a practical example of considerable physical
applicability. In much of modern physics a bosonic mode is coupled
to a reservoir, with which it can exchange particles. The traditional
example of this is a single-mode interferometer\cite{Louisell}. In
current applications relevant to quantum information, one may have
a localized photonic waveguide mode\cite{Optical_microcavities_Vahala},
an ultra-cold Bose condensate\cite{Inguscio_Nature_BEC2008}, or a
nano-mechanical oscillator\cite{Painter_Nature_Optomechanical}. These
exchange photons, atoms or phonons respectively with their environments.
In nearly all of these recent applications, one is interested in evolution
with non-Markovian reservoirs.

In order to model such physically important examples and evaluate
the coarse grained entropy, we consider the following non-Markovian
system. A set of bosonic modes (the\emph{ system}) is coupled to a
large number of other modes (the \emph{reservoir}). The total Hamiltonian
can be written, on introducing $\hat{n}_{ij}=\hat{a}_{i}^{\dagger}\hat{a}_{j}$,
as:

\begin{equation}
\hat{H}=\hbar\sum_{ik}\omega_{ik}\hat{n}_{ik},\label{eq:H_bose_reservoir-1}
\end{equation}
where $\omega_{ik}=\delta_{ik}\nu_{k}+g_{ik}$. Here $\nu_{k}$ denotes
the resonant frequencies of the modes, $g_{kj}$ denotes the couplings
between the modes. We assume that the modes for $k=1,\dots S$ are
system modes ($A$), while the remainder are the reservoir (B). We
note that we make no assumptions concerning their relative sizes or
quantum states. We suppose that the initial density matrix at time
$t=0$ has the general number-conserving form:

\begin{equation}
\widehat{\rho}^{0}=\int P^{0}(\bm{\underline{n}})\Lambda(\underline{\bm{n}})d\underline{\bm{n}}\label{eq:dens_matrix_t0}
\end{equation}

The Gaussian representation provides a form to express the real or
imaginary time evolution of the density matrix of either fermions
and bosons into a set of phase-space stochastic equations \cite{Corney:2003,Corney2004a,corney_drummond_06c},
the mappings are given by \cite{Corney:2003,corney_drummond_06c}:

\begin{eqnarray}
\hat{n}_{ik}\hat{\rho} & \rightarrow & \left[n_{ik}-\frac{\partial}{\partial n_{\ell m}}(1\pm n_{im})n_{\ell k}\right]P,\nonumber \\
\hat{\rho}\hat{n}_{ik} & \rightarrow & \left[n_{ik}-\frac{\partial}{\partial n_{\ell m}}n_{im}(1\pm n_{\ell k})\right]P,\nonumber \\
\label{eq:mappings_bf}
\end{eqnarray}
Here the +(-) sign corresponds to the bosonic (fermionic) case respectively,
and we will use the bosonic identities here.

The real time evolution of the density matrix is given, as usual,
by:

\begin{equation}
\frac{d\hat{\rho}}{dt}=-\frac{i}{\hbar}[\hat{H},\,\hat{\rho}].\label{eq:ev_dens_op}
\end{equation}
Using Eq.$\ $(\ref{eq:ev_dens_op}) and Eq.$\ $(\ref{eq:mappings_bf})
we obtain that the time evolution equation of the stochastic Green's
function in matrix form is:

\begin{equation}
\underline{\overset{\centerdot}{\bm{n}}}=i\left[\underline{\bm{\omega}},\,\underline{\bm{n}}\right].\label{eq:time_ev_nbose}
\end{equation}

The solution of Eq.$\ $(\ref{eq:time_ev_nbose}) is simply: 

\begin{equation}
\underline{\bm{n}}(t)=e^{i\underline{\bm{\omega}}t}\underline{\bm{n}}(0)e^{-i\underline{\bm{\omega}}t}.\label{eq:nt_bose}
\end{equation}
The reduced linear entropy for this system is evaluated using Eq.$\ $(\ref{eq:S2_red})
and the result for the time evolution of the stochastic Green's functions,
Eq.$\ $(\ref{eq:nt_bose}). In this case we will trace over the system
$A$. Therefore, the expression for the reduced entropy is:

\begin{eqnarray}
S_{2}^{red} & = & -\ln\iint P^{0}(\underline{\bm{n}})P^{0}(\underline{\bm{n}}')\det[1+\underline{\bm{n}}^{A}(t)+\underline{\bm{n}}'^{A}(t)]^{-1}d\underline{\bm{n}}d\underline{\bm{n}}'.\nonumber \\
\label{eq:S2b1}
\end{eqnarray}

\subsection{Example of bosonic entropy}

In order to illustrate the time evolution of the coarse grained entropy
of Eq. (\ref{eq:S2b1}), we will consider the following model for
the Hamiltonian of Eq. (\ref{eq:H_bose_reservoir-1}). The bosonic
system will be a single mode ($j=1$) and the reservoir will be modeled
as a Lorentzian distribution of couplings, with:

\begin{equation}
g_{j}=\frac{C}{\nu_{j}^{2}+s^{2}},\label{eq:g}
\end{equation}
where C is the strength of the coupling, $\nu_{j}=\pm jd\omega$ are
the resonant frequencies of the modes, and $s$ describes the non-Markovian
reservoir width. For this model, the time evolution of the stochastic
Green's function of Eq. (\ref{eq:nt_bose}) is written assuming that
$\underline{\bm{n}}(0)$ describes the thermal state at $t=0$, with
a system occupation of $n=1$ , and all the other modes unoccupied.
Here $\underline{\bm{\omega}}$ is an $M\times M$-matrix, and $M$
is total the number of modes, so that:

\begin{equation}
\underline{\bm{\omega}}=\left(\begin{array}{cccccc}
\nu_{11} & g_{21} & \cdots & \cdots & \cdots & g_{1M}\\
g_{12} & \ddots & 0 & \cdots & 0 & \vdots\\
\vdots & 0 & \ddots & 0 & \vdots & \vdots\\
\vdots & \vdots & \ddots & \ddots & 0 & \vdots\\
\vdots & 0 & \cdots & 0 & \ddots\\
g_{M1} & \cdots & \cdots & \cdots &  & \nu_{MM}
\end{array}\right).\label{eq:WM}
\end{equation}

\begin{figure}
\includegraphics[scale=0.45]{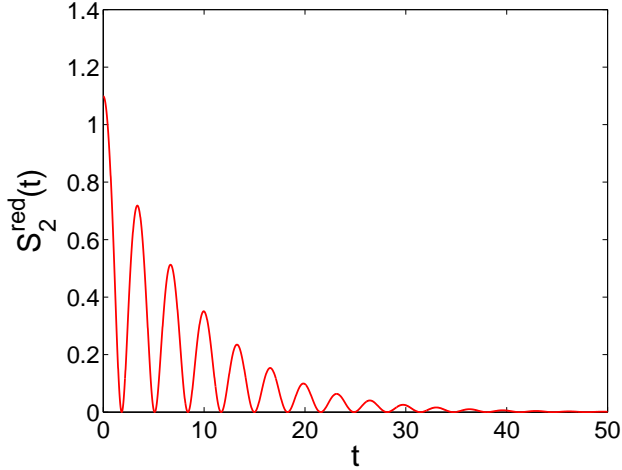}

\caption{Time evolution of the coarse grained entropy for a bosonic thermal
state coupled to a non-Markovian reservoir. \label{fig:St_bosons}}

\end{figure}

In Figure (\ref{fig:St_bosons}), we show the time evolution of the
coarse grained entropy. We use the following parameters: $M=100$,
$s=0.5$, $C=0.05$. The frequency spacing of the modes is $d\omega=0.04$.
We observe that the non-Markovian behavior of the reservoir, as shown
in the increase and decrease of the entropy with time. 

In summary, to solve for the coarse grained entropy one must take
the partial determinant of a block reduced form $\underline{\bm{n}}^{A}$
of the time-evolved stochastic Green's function, then average over
the initial ensemble. Apart from the limitation to linear couplings
needed to obtain an exactly soluble form, there are no restrictions
to the state, the type of coupling or the subdivision between the
system and the reservoir in this calculation.

\section{Gaussian representations for fermions}

The fermionic case is similar, except that one must use fermionic
coherent states\cite{Cahill1999a} and Grassmann integrals to carry
out the trace calculations. Just as with bosons, this has a clear
intuitive meaning. In this approach, the physical many-body system
is treated as a distribution over fermionic Green's functions, whose
average in the observed Green's function or correlation function.
As with the bosonic case, the stochastic Green's functions themselves
can be nonhermitian.

\subsection{Un-normalized Gaussians}

Here we must evaluate the trace of the product of two un-normalized
fermionic Gaussian operators, $F\left(\underline{\bm{\mu}},\underline{\bm{\nu}}\right)=Tr\left[\hat{\Lambda}_{u}\left(\underline{\bm{\mu}}\right)\hat{\Lambda}_{u}\left(\underline{\bm{\nu}}\right)\right]$
for the M-mode case:

\begin{equation}
F\left(\underline{\bm{\mu}},\underline{\bm{\nu}}\right)=Tr\left[:e^{-\hat{\bm{a}}^{\dagger}\underline{\bm{\mu}}\hat{\bm{a}}}::e^{-\hat{\bm{a}}^{\dagger}\underline{\bm{\nu}}\hat{\bm{a}}}:\right],
\end{equation}

For fermions\cite{Cahill1999a}, the trace of an operator using fermionic
coherent states $\vert\bm{\alpha}\rangle$ in terms of Grassmann variables
$\bm{\alpha}$ is:

\begin{equation}
Tr[\hat{O}]=\int d^{2M}\bm{\alpha}\langle-\bm{\alpha}\vert\hat{O}\vert\bm{\alpha}\rangle,
\end{equation}
 and the identity operator is:

\begin{equation}
\int d^{2M}\bm{\alpha}\vert\bm{\alpha}\rangle\langle\bm{\alpha}\vert=1.
\end{equation}

Therefore, we have: 
\begin{eqnarray}
F\left(\underline{\bm{\mu}},\underline{\bm{\nu}}\right) & = & \frac{1}{\pi^{2M}}\int d^{2M}\bm{\alpha}d^{2M}\bm{\beta}\left\langle -\bm{\alpha}\right|:e^{-\hat{\bm{a}}^{\dagger}\underline{\bm{\mu}}\hat{\bm{a}}}:\times\nonumber \\
 &  & \times\left|\bm{\beta}\right\rangle \left\langle \bm{\beta}\right|:e^{-\hat{\bm{a}}^{\dagger}\underline{\bm{\nu}}\hat{\bm{a}}}:\left|\bm{\alpha}\right\rangle .
\end{eqnarray}
 Expanding the normal-ordered exponential, and using the standard
eigenvalue properties for the fermionic coherent states: $\hat{\bm{a}}\left|\bm{\alpha}\right\rangle =\bm{\alpha}\left|\bm{\alpha}\right\rangle $
gives:

\begin{eqnarray}
F\left(\underline{\bm{\mu}},\underline{\bm{\nu}}\right) & = & \int d^{2M}\bm{\alpha}d^{2M}\bm{\beta}\left\langle -\bm{\alpha}\right|e^{\bm{\alpha}^{\dagger}\underline{\bm{\mu}}\bm{\beta}}\times\nonumber \\
 &  & \left|\bm{\beta}\right\rangle \left\langle \bm{\beta}\right|e^{-\bm{\beta}^{\dagger}\underline{\bm{\nu}}\bm{\alpha}}\left|\bm{\alpha}\right\rangle .
\end{eqnarray}
 From the inner product of two fermion coherent states, we note that:
\begin{equation}
\left\langle \bm{\alpha}\right|\bm{\beta}\rangle=e^{\bm{\alpha}^{\dagger}\bm{\beta}-\left(\bm{\alpha}^{\dagger}\bm{\alpha}+\bm{\beta}^{\dagger}\bm{\beta}\right)/2}\,,
\end{equation}

Next, introducing a double-dimension Grassmann vector : 
\begin{equation}
\bm{\gamma}=\left[\begin{array}{c}
\bm{\alpha}\\
\bm{\beta}
\end{array}\right]\,,
\end{equation}
we finally obtain a Gaussian Grassmann integral over $2M$ complex
coordinates, which we can write this as: 
\begin{eqnarray}
F\left(\underline{\bm{\mu}},\underline{\bm{\nu}}\right) & = & \int d^{4M}\bm{\gamma}e^{\bm{\alpha}^{\dagger}\underline{\bm{\mu}}\bm{\beta}-\bm{\beta}^{\dagger}\underline{\bm{\nu}}\bm{\alpha}-\bm{\alpha}^{\dagger}\bm{\beta}+\bm{\beta}^{\dagger}\bm{\alpha}^{\dagger}-\left(\bm{\alpha}^{\dagger}\bm{\alpha}+\bm{\beta}^{\dagger}\bm{\beta}\right)}\nonumber \\
 & = & \int d^{4M}\bm{\gamma}e^{-\bm{\gamma}^{\dagger}\underline{\bm{\Gamma}}\bm{\gamma}}\nonumber \\
 & = & \det\left[\underline{\bm{\Gamma}}\right]
\end{eqnarray}
 Here we have used the standard identity(\cite{Integrals}) for an
N-dimensional Gaussian complex Grassmann integrals, and introduced
a double-dimension matrix, 
\begin{equation}
\underline{\bm{\Gamma}}=\left[\begin{array}{cc}
\underline{\bm{I}} & \underline{\bm{I}}-\underline{\bm{\mu}}\\
\underline{\bm{\nu}}-\underline{\bm{I}} & \underline{\bm{I}}
\end{array}\right]
\end{equation}

Therefore, on simplifying the determinant, we obtain: 
\begin{equation}
F\left(\underline{\bm{\mu}},\underline{\bm{\nu}}\right)=\det\left[\bm{\mathrm{I}}\,+\left(\underline{\bm{I}}-\underline{\bm{\mu}}\right)\left(\underline{\bm{I}}-\underline{\bm{\nu}}\right)\right]
\end{equation}

\subsection{Normalized Gaussians}

Just as in the bosonic case, it is useful to rewrite these expressions
in terms of the normally ordered Green's functions or correlations
of the basis sets, defined so that: 
\begin{equation}
n_{ij}=Tr\left[\widehat{\Lambda}(\mathbf{n})\hat{a}_{i}^{\dagger}\hat{a}_{j}\right]
\end{equation}
 Here, introducing the hole Green's functions, $\tilde{\underline{\mathbf{n}}}=\left[\underline{\bm{\mathrm{I}}}-\underline{\mathbf{n}}\right]$,
and $\tilde{\underline{\mathbf{m}}}=\left[\underline{\bm{\mathrm{I}}}-\underline{\mathbf{m}}\right]$,
the normalized generators are: 
\begin{equation}
\widehat{\Lambda}(\mathbf{n})=\Omega\det\left[\tilde{\underline{\mathbf{n}}}\right]:\exp\left[\hat{\bm{a}}^{\dagger}\left[\tilde{\underline{\mathbf{n}}}^{-1}-2\underline{\bm{\mathrm{I}}}\right]^{T}\hat{\bm{a}}\right]:
\end{equation}
 and therefore, 
\begin{eqnarray}
\underline{\bm{\nu}}^{T} & = & 2\underline{\bm{\mathrm{I}}}-\tilde{\underline{\mathbf{n}}}^{-1}\\
\underline{\bm{\mu}}^{T} & = & 2\underline{\bm{\mathrm{I}}}-\tilde{\underline{\mathbf{m}}}^{-1}
\end{eqnarray}
 Hence, 
\begin{equation}
\tilde{\underline{\mathbf{n}}}\left(\underline{\bm{I}}-\underline{\bm{\nu}}\right)^{T}=\underline{\mathbf{n}}
\end{equation}
which leads to the following result for the normalized inner product:
\begin{equation}
Tr\left[\widehat{\Lambda}(\mathbf{m})\widehat{\Lambda}(\mathbf{n})\right]=\det\left[\tilde{\underline{\mathbf{n}}}\tilde{\underline{\mathbf{m}}}+\underline{\mathbf{n}}\underline{\mathbf{m}}\right]\label{eq:Tracef}
\end{equation}

We note that this has some obvious properties. Suppose that $\underline{\mathbf{n}}$
and $\underline{\mathbf{m}}$ are each diagonal in the same basis,
with real eigenvalues $n_{i}$ such that $0\leq n_{i}\leq1$ . Then
one obtains:
\begin{equation}
Tr\left[\widehat{\Lambda}(\mathbf{m})\widehat{\Lambda}(\mathbf{n})\right]=\prod_{i=1}^{M}\left(\tilde{n}_{i}\tilde{m}_{i}+n_{i}m_{i}\right)
\end{equation}
Thus, the two generators are orthogonal if, in any mode, one generator
has a vanishing particle population while the other has a vanishing
hole population. The overlap is maximized if the generators both have
a unit hole population or a unit particle population in all modes. 

For the thermal case, the entropy can be evaluated in other ways,
but here we demonstrate the technique using the Gaussian operator
method, which will be useful to evaluate the entropy of other systems.

\subsection{Thermal linear entropy for fermions}

We can now apply these inner products to the evaluation of the linear
entropy of a thermal Fermi-Dirac states. When the density matrix is
thermal, only a single basis set member is required, and:
\begin{eqnarray}
S_{2} & = & -\ln Tr\left(\widehat{\Lambda}^{2}(\mathbf{n}_{th})\right)\nonumber \\
 & = & -\ln\det\left[\underline{\bm{\mathrm{I}}}-2\underline{\mathbf{n}}_{th}+2\underline{\mathbf{n}}_{th}^{2}\right]
\end{eqnarray}
Just as with bosons, for the thermal case we know that the thermal
Green's function is a scalar: $\underline{\mathbf{n}}_{th}=n_{th}$,
where $n_{th}$ is now the Fermi-Dirac occupation number at a given
temperature, so that $0\leq n_{th}\leq1$. Here the results are asymptotically
equal to the bosonic case as expected for $n_{th}\ll1$ or for $\tilde{n}_{th}\ll1$
. Typical results are shown in Table (II), showing the particle-hole
symmetry. The greatest entropy is at $n_{th}=0.5$, corresponding
to infinite temperature, while mirror states with small hole occupations
can be thought of as having negative temperatures or negative Hamiltonians.

\begin{table}
\begin{tabular}{|c|c|}
\hline 
$n_{th}$ & $S_{2}$\tabularnewline
\hline 
0.01 & 0.02\tabularnewline
\hline 
0.1 & 0.1984\tabularnewline
\hline 
0.5 & 0.6931\tabularnewline
\hline 
0.9 & 0.1984\tabularnewline
\hline 
0.99 & 0.02\tabularnewline
\hline 
\end{tabular}

\caption{Linear entropy for thermal states using the Gaussian representation
for fermions.}
\label{Table:Linear_entropy_GRB-1}
\end{table}

\subsection{Coarse grained entropy for fermions.}

Just as in the case of bosons, we now consider an example of physical
applicability for the case of fermions that is a fermionic mode coupled
to a reservoir. An example of such a system is a quantum dot coupled
to a fermionic reservoir \cite{DalgarnoQuantumDotPhysRevLett.100.176801},
or a fermionic atom-tronic circuit\cite{DasAubinAtomtronicsPhysRevLett.103.123007}.
This system can be considered as an example of solid-state quantum
physics and has potential applications in quantum information processing.
In such hybrid quantum systems, long-range interactions can be important.
Here we neglect this in order to obtain analytic results, although
these can be added if necessary.

We consider an identical Hamiltonian to the bosonic case in the last
section. Similar to the bosonic case, we assume that the modes for
$k=1,\ldots S$ are system modes $(A)$, while the remainder are for
the reservoir $(B)$ and the initial density matrix at time $t=0$
has the general number-conserving of Eq.$\ $(\ref{eq:dens_matrix_t0}).

Using the identities of Eq.$\ $(\ref{eq:mappings_bf}) and Eq.$\ $(\ref{eq:ev_dens_op})
we obtain the time evolution equation of the stochastic Green's function
is identical to the bosonic case. Therefore, the expression for the
reduced linear entropy for fermions is:

\begin{equation}
S_{2}^{red}=-\ln\iint P^{0}(\underline{\bm{n}})P^{0}(\underline{\bm{n}}')\det[\tilde{\underline{\bm{n}}}^{A}(t)\underline{\tilde{\bm{n}}}'^{A}(t)+\underline{\bm{n}^{A}}(t)\underline{\bm{n}}'^{A}(t)]d\underline{\bm{n}}d\underline{\bm{n}}'.\label{eq:St_fermions}
\end{equation}

This has a very simple physical interpretation. The linear entropy
and hence the fermionic entanglement can be calculated completely
from the local stochastic Green's functions in the system of interest.
However, one must average over all possible initial states defined
by the complete initial phase-space distribution $P^{0}(\underline{\bm{n}})$.
This is necessary, since the correlations and initial states of the
reservoir can change the final system properties.

\subsection{Example of fermionic entanglement}

Similar to the bosonic case, we will illustrate the time evolution
of the coarse grained entropy of Eq. (\ref{eq:St_fermions}). We will
consider a pure fermionic number state. The $\underline{\bm{\omega}}$
matrix of time evolution of the stochastic Green's function has the
same form as the bosonic case, Eq. (\ref{eq:WM}). We also model the
reservoir with a Lorentzian distribution described by Eq. (\ref{eq:g}). 

\begin{figure}
\includegraphics[scale=0.45]{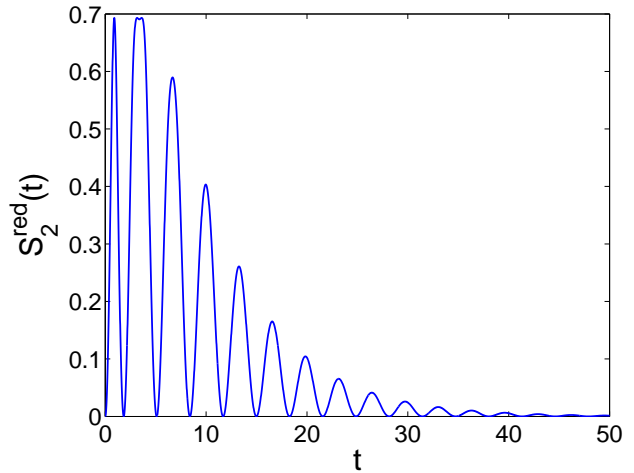}

\caption{Time evolution of the coarse grained entropy using the fermionic Gaussian
representation, for a pure fermionic number state coupled to a non-Markovian
reservoir.\label{fig:St_fermions}}
\end{figure}

In Figure$\ $(\ref{fig:St_fermions}), we show the time evolution
of the coarse grained entropy, Eq.$\ $(\ref{eq:St_fermions}), using
an identical model and parameters to the bosonic case. However, there
is a large physical difference, as a fermionic state with $n=1$ is
a pure number state. We observe an initial increase of entropy, which
means that the entropy is measuring the entanglement of the system
with the reservoir. As before, the non-Markovian behavior of the system
is shown in the increase and decrease of the coarse-grained entropy,
which in this case corresponds to entanglement oscillations. 

\section{Summary}

We have calculated the linear entropy for sampled phase-space representations
of bosonic and fermionic quantum many-body systems. The crucial element
to the calculation is an evaluation of the inner products of the phase-space
basis elements. Traditional displacement-based phase-space methods
have a range of pathologies. In the Wigner and Husimi cases, the inner
products are singular or divergent, while in the Glauber-Sudarshan
case, the representation is not well-defined in all cases. Even the
positive-P distribution, which exists and has well-defined inner products,
we find there is a sampling convergence problem. By comparison, Gaussian
phase-space representations for fermions and bosons are much more
suitable for the task. For thermal states, only a single basis element
is needed, and the inner-products are well-behaved. 

There is a counter-intuitive element to the idea that entropy is conserved
in quantum dynamics; but this must be the case when simulating time-reversible,
unitary quantum dynamics. We show how, in the case of reduced entropy
of a subsystem, the linear entropy can and does evolve in time. We
give an exact calculation of couplings of Fermi and Bose systems to
a non-Markovian quantum reservoir. Such phase-space methods appear
useful for investigating the fundamental paradox of entropy-invariance
in unitary quantum dynamics. They are equally applicable to entire
system evolution and to the evolution of the density matrix for a
coarse-grained sub-space.
\begin{acknowledgments}
L. E. C. R. Z. acknowledges financial support from CONACYT-Mexico.
PDD acknowledges financial support from the Australian Research Council
and Swinburne University of Technology, as well as the generous hospitality
of the Aspen Center for Physics.
\end{acknowledgments}
\bibliographystyle{apsrev4-1}
\bibliography{EntropyRefs/entropy_ref,EntropyRefs/phase-space,EntropyRefs/Renyi_references,EntropyRefs/REf_first_r,EntropyRefs/non_markovian}

\end{document}